\documentclass[conference]{IEEEtran}
\IEEEoverridecommandlockouts

\usepackage{latexsym}
\usepackage{graphicx}
\usepackage{amsfonts,amssymb,amsmath}
\usepackage{hyperref}
\usepackage{subcaption}
\usepackage{siunitx}
\usepackage{amsmath,amssymb,amsfonts}
\usepackage{algorithmic}
\usepackage{algorithm}
\usepackage{array}

\usepackage{textcomp}
\usepackage{url}
\usepackage{graphicx}

\usepackage{tikz}
\usetikzlibrary{arrows.meta,positioning,fit,calc,shapes.geometric}
\usepackage{threeparttable}
\usepackage{multirow}
\usepackage{cite}
\usepackage{booktabs}
\usepackage{xcolor}
\usepackage{epsfig}
\usepackage{epstopdf}
\usepackage{acronym}
\graphicspath{{fig/}}
\usepackage{mathtools}
\usepackage{amsthm}
\usepackage{bm}
\usepackage{comment}

  \usepackage{graphicx}
  \usepackage{lipsum}
  \usepackage[absolute,overlay]{textpos}

\usepackage{cleveref}

\usepackage{bbm}

\newcommand{\subsf}{\sf \scriptscriptstyle}
\newcommand{\bs}{\boldsymbol}

\newcommand{\wh}{\widehat}

\ifCLASSINFOpdf

\else

\fi

\usepackage{algorithmic}
\usepackage{graphicx}
\usepackage{textcomp}
\usepackage{xcolor}
\def\BibTeX{{\rm B\kern-.05em{\sc i\kern-.025em b}\kern-.08em
    T\kern-.1667em\lower.7ex\hbox{E}\kern-.125emX}}

\begin{document}

\title{Transformer-Based Rate Prediction for Multi-Band Cellular Handsets
}

\author{
 \IEEEauthorblockN{Ruibin Chen$^{\ast1}$, Haozhe Lei$^{\ast\dagger1}$, Hao Guo$^{1,2}$, \\ Marco Mezzavilla$^3$, Hitesh Poddar$^4$, Tomoki Yoshimura$^4$, Sundeep Rangan$^1$}\\
 \IEEEauthorblockA{$^1$NYU WIRELESS, Tandon School of Engineering, New York University, Brooklyn, NY 11201, USA\\
 $^2$Department of Electrical Engineering, Chalmers University of Technology, Gothenburg, Sweden\\
 $^3$Dipartimento di Elettronica, Informazione e Bioingegneria (DEIB), Politecnico di Milano, Milan, Italy\\
 $^4$Sharp Laboratories of America (SLA), Vancouver, Washington, USA\\
 \{rc5018, hl4155, hg2891, srangan\}@nyu.edu; marco.mezzavilla@polimi.it; \{poddarh, yoshimurat\}@sharplabs.com}
}

%
%

\markboth{Journal of \LaTeX\ Class Files,~Vol.~14, No.~8, August~2015}%
{Shell \MakeLowercase{\textit{et al.}}: Bare Demo of IEEEtran.cls for IEEE Journals}
%



\maketitle
\begingroup
\renewcommand\thefootnote{}
\footnotetext{$^{\ast}$ These authors contributed equally to this work.}
\footnotetext{$^{\dagger}$ Corresponding author: Haozhe Lei (hl4155@nyu.edu).}
\endgroup

\begin{abstract}
Cellular wireless systems are facing a proliferation of frequency bands over a wide spectrum, particularly with the expansion into FR3. These bands must be supported in user equipment (UE) handsets with multiple antennas in a constrained form factor. Rapid variations in channel quality across the bands from motion and hand blockage, limited field-of-view of antennas, and hardware and power-constrained measurement sparsity pose significant challenges to reliable multi-band channel tracking. This paper formulates the problem of predicting achievable rates across multiple antenna arrays and bands with sparse historical measurements. We propose a transformer-based neural architecture that takes asynchronous rate histories as input and outputs per-array rate predictions. Evaluated on ray-traced simulations in a dense urban micro-cellular setting with FR1 and FR3 arrays, our method demonstrates superior performance over baseline predictors, enabling more informed band selection under realistic mobility and hardware constraints.
\end{abstract}


%

\section{Introduction}


Modern commercial handsets are increasingly required to support a growing number of frequency bands with multiple distinct antennas in a compact form factor \cite{wang2024review}.  Continuous and reliable coverage requires monitoring these bands
and initiating frequent handover and serving cell decisions as the UEs move in the environment. 
There are several factors in emerging systems that are now making this multi-band channel tracking challenging.

\begin{figure}[t]
    \centering
    \includegraphics[width=0.9\linewidth]{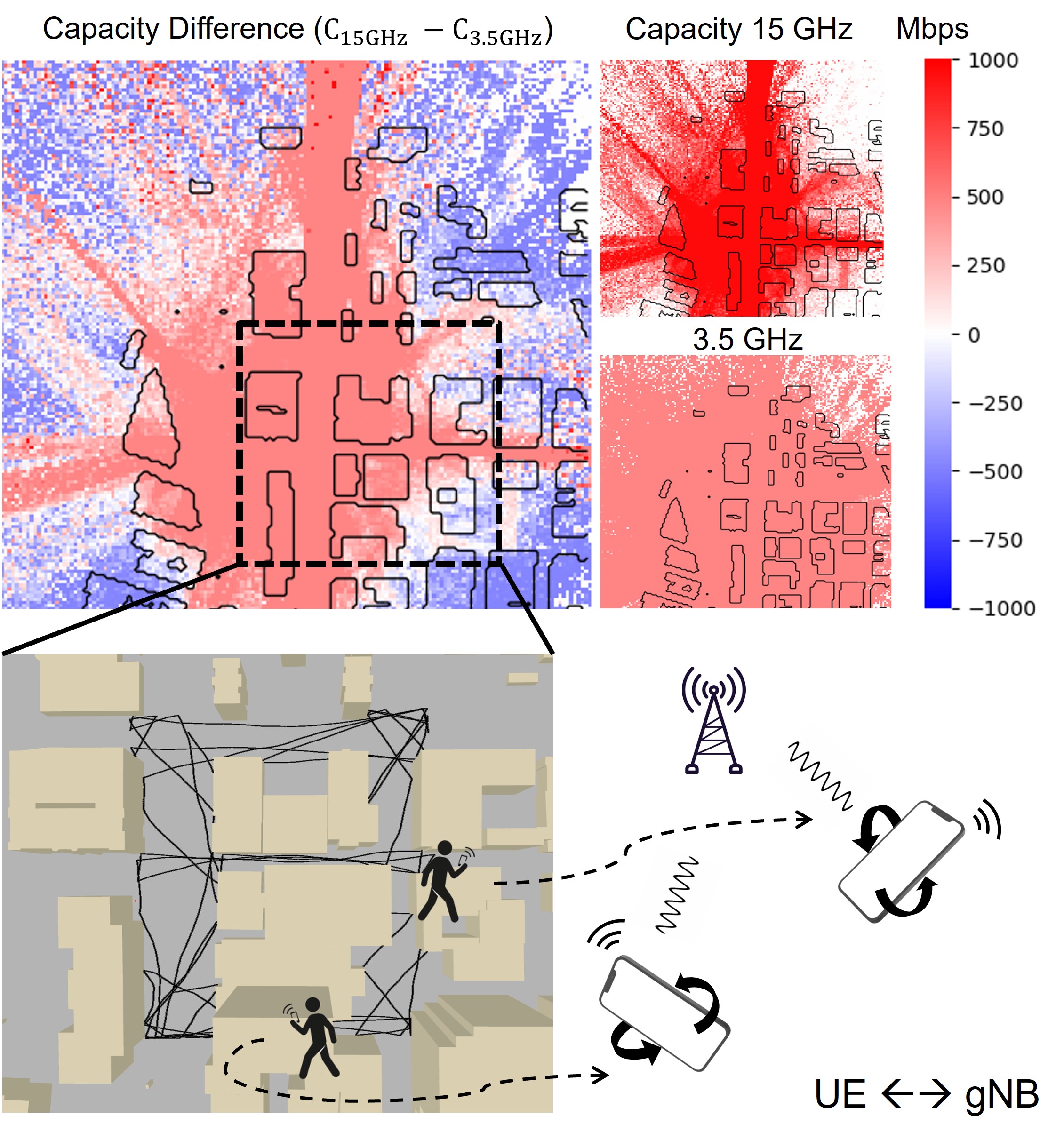}
    \caption{Top: Ray tracing capacity in an urban micro–cell. Top-left: difference $\Delta C=C_{15}-C_{3.5}$ (red$>$0$\rightarrow$\SI{15}{GHz}, blue$<$0$\rightarrow$\SI{3.5}{GHz}, Mbps). Top-right: per-band capacity at \SI{15}{GHz} and \SI{3.5}{GHz} (shared color bar, Mbps). Bottom: zoom of the dashed area with the site model and simulated pedestrian trajectories with handset rotations, for UE mobility and UE–gNB interaction, yielding time-varying band/array rates.}
    \label{fig:intro}
\end{figure}

\textbf{Proliferation of bands with rapidly varying coverage}:  Most importantly, the number of frequency bands is exploding -- for example, 3GPP identifies more than 100 bands in FR1 alone \cite{3gpp38.101}.  The number of bands is likely to accelerate with the expansion into FR3 \cite{kang2024cellular,testolina2024sharing,shakya2024propagation}.  The coverage from each of these bands can vary rapidly over space.  For example, the top of ~\Cref{fig:intro} shows the predicted capacity (or Rate) based on ray tracing at two bands -- \SI{3.5}{GHz} and \SI{15}{GHz} in an urban micro-cellular setting.  Details of the simulation are presented below.  We observe that the regions where one band is better than the other vary rapidly due to complex propagation differences between the bands.  This fact necessitates that UEs will need to switch bands rapidly to maintain optimal throughput. 

\textbf{UE antenna constraints}:  Many bands may require
distinct antennas, yet the available real estate within handsets is severely limited  \cite{wang2024review}.  Moreover, individual antennas often exhibit narrow steerable ranges and single polarization, implying that a coverage for any given band will have a limited field of view.  Coverage can also be limited by hand blocking
~\cite{Raghavan18CommMag,Raghavan22TAP}.  As a result, natural changes in the orientation of the phone will necessitate further rapid band switching.  

\textbf{Limited measurements}:  To reduce cost and energy consumption,
multiple antennas may be switched to a single RF chain.
In addition, it may be too costly for the power to operate and monitor multiple bands simultaneously.
Consequently, UEs are forced to make band selection decisions on limited measurements.



\textbf{Problem}:  Motivated by these developments, 
we formulate a general problem of tracking and predicting channel rates from multiple antenna arrays.  Different arrays may be associated with different frequency bands.  For each array, the UE has limited prior measurements of the available rate since the array may have been switched off to save power, or the RF chain for that array may have been switched to a different array.  
The problem is to estimate the achievable rate on each array and to predict which array offers the highest rate.

\textbf{Our contributions}:  For this problem, 
we propose a novel transformer neural network architecture
that takes as inputs the past measured channel quality indicator (CQI) on each link, a link being a potential connection from a serving cell to a particular UE array in some carrier frequency.
The measured CQIs may have been at arbitrary time points in the past.  This multi-dimensional CQI history is fed to a temporal convolutional encoder followed by a single transformer layer.

The proposed method is simulated on a realistic antenna
layout with two antennas in FR1 (\SI{3.5}{GHz}) and
two antenna arrays in FR3 (\SI{15}{GHz}).  Ray tracing simulation is used to estimate rates across the antenna arrays with both UE orientation and translational motion in a dense urban setting.  It is shown that the method outperforms simple baselines.

\subsection*{Related Work}
Measurement-driven studies with form-factor UEs at \SI{28}{GHz} show that hand/body occlusion and device rotation dominate coverage and beam management, challenging simplified blockage models~\cite{Raghavan18CommMag,Raghavan22TAP}. Multi-band campaigns (11/16/28/32\,GHz) quantify $\sim$20\,dB human-blockage loss and validate KED/UTD abstractions~\cite{Qi17IWCMC,Virk20TAP}, while FR1/FR3 results at 6.75/16.95\,GHz provide calibrated outdoor models and indoor-hotspot angular-spread statistics~\cite{Shakya24Outdoor,Ying24AngularSpread}.

EM/RT pipelines benchmark KED/UTD/PO against 60\, GHz measurements to expose accuracy–complexity trade-offs~\cite{Mukherjee22Access}; device-as-scatterer CAD reveals panel-dependent pattern loss and reflection beyond fixed-angle masks~\cite{Fernandes22Access}. Site-specific calibration/validation of {NYURay} at upper mid-band tightens agreement with measurements, and {NYUSIM} offers spatially consistent, blockage-aware corpora for synthesis and analysis~\cite{Ying24RayTracing,Ju19NYUSIM}.

Predictors enable proactive switching and lightweight channel inference: AE+LSTM sequence models improve Top-$C$ ranking and reduce misalignment for next-slot beam/BS prediction~\cite{Shah22OJCOMS}; cross-band learning maps sub-6\, GHz features to mmWave beam/blockage; vision/semantics aid LoS/NLoS forecasting and candidate selection~\cite{Alrabeiah20TCOM,Xu23TWC,Yang23Semantics}. Physics-informed Reinforcement Learning (RL) further embeds channel dynamics, demonstrating zero-shot and DT-in-the-loop gains with growing industrial traction~\cite{LeiICRA2024,Lei2025DTWIN,Li25RLPhysics,Lei2025_FullPosteriorLocalization}.

\section{Frequency Hopping Problem Formulation}

Consider a UE that can be served in the downlink from one of $N_{\subsf BS}$ base station (gNB) cells
with reception on one of $N_{\subsf arr}$ antenna arrays.
We define a \emph{link} as a pair $(i,j)$ from gNB cell $i$ to antenna array $j$, and we assume there are $M$
potential links.  Each link $i=1,\ldots,M$ is associated with some carrier frequency $f^{(i)}$.
Two different links may use the same frequency or the same serving cell.  Also, depending on the bandwidth of the antenna,
two different serving cells could be received in the same antenna array.

\begin{figure}[t]
    \centering
    \includegraphics[width=1\linewidth]{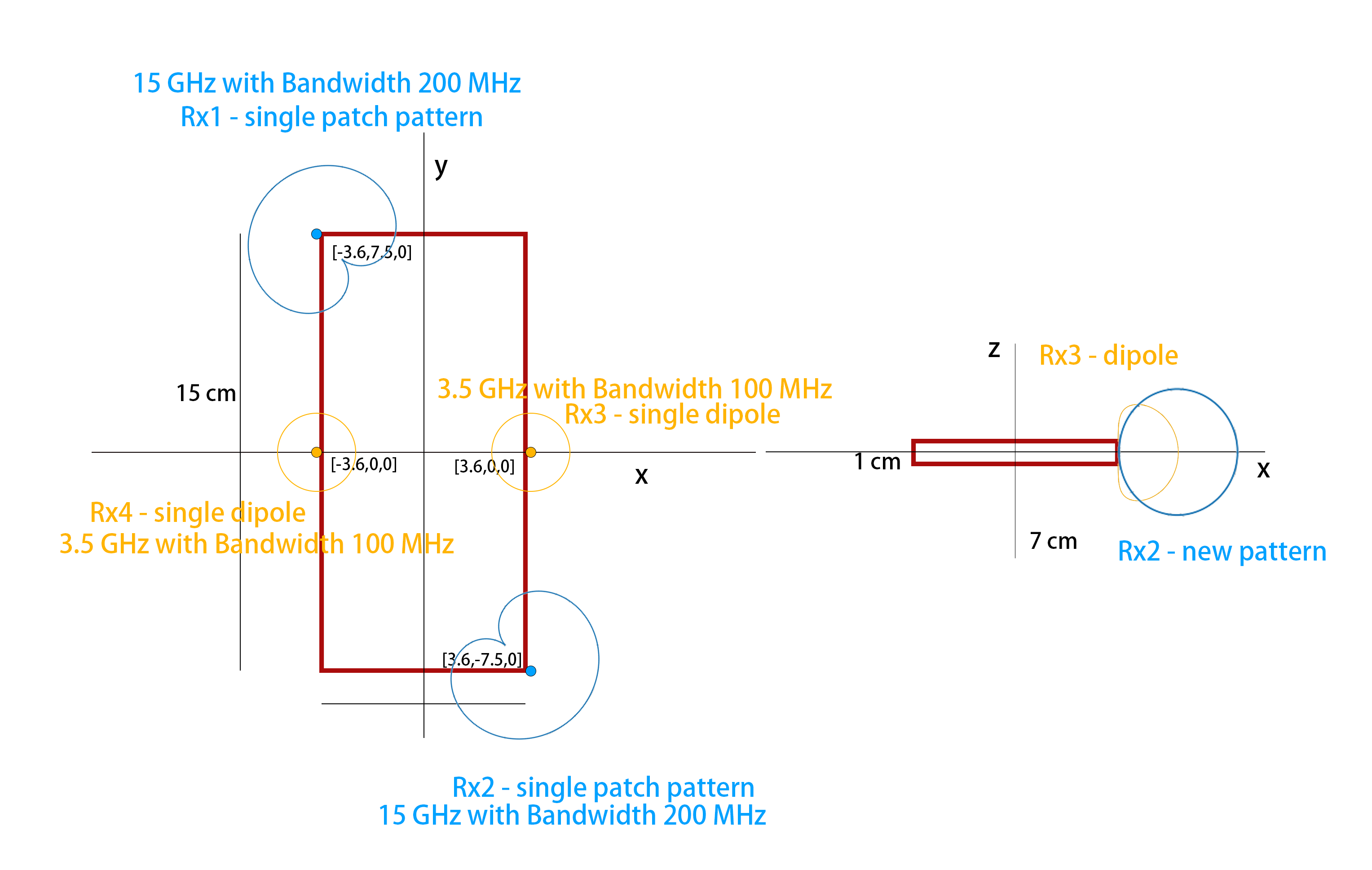}
    \caption{Illustration of UE radian pattern. The red rectangular shows the UE body shape as 7cm $\times$ 15cm $\times$ 1cm. New directive 15GHz FR3 radian pattern is shown in blue while the traditional 3.5GHz dippole pattern is shown as in yellow. The figure demostrates the horizontal cut of the UE radian pattern in left with virtical cut in right.}
    \label{fig:ue_radian}
\end{figure}

As one simple example that we will use in the simulations below, we consider a UE being served by $N_{\subsf BS}=2$ 
potential cells that are co-located, so they are from a single gNB site.  
The two cells operate in two distinct frequencies:
BS1 at \SI{3.5}{GHz} in FR1 and BS2 \SI{15}{GHz} in FR3 (details in \Cref{table:simulation}.).  The UE antenna layout that we will use in the simulations below is  
\Cref{fig:ue_radian}.  There are $N_{\subsf arr}=4$ antennas 
-- two dipole antennas  at \SI{3.5}{GHz} and two patch antennas at 
at \SI{15}{GHz}.  Four antennas are labeled as RX1 to RX4.
In this case, there are $M=4$ potential links:
\begin{itemize}
    \item Link 1:  From BS1 to RX3 at $f^{(1)}=$\,\SI{3.5}{GHz}
    \item Link 2:  From BS1 to RX4 at $f^{(2)}=$\,\SI{3.5}{GHz}
    \item Link 3:  From BS2 to RX1 at $f^{(3)}=$\,\SI{15}{GHz}
    \item Link 4:  From BS2 to RX2 at $f^{(4)}=$\,\SI{15}{GHz}
\end{itemize}
Each antenna has a directivity pattern, so the link quality from the same serving cell may be different for two different
antennas.  For example, the link quality on link 1 may be different than link 2 depending on the orientation of
the different antennas relative to the angles of arrivals of the paths.  
In \Cref{fig:ue_radian}, the antenna placement, orientation, and antenna pattern follow the 3GPP simulation model \cite{3gpp_release19}.  
In the simulations below, 
we will also consider blockage from the metal UE body itself that influences coverage depending on orientation.


Our broad problem is to track the link quality on the $M$ links.  
Time is divided into measurement periods of length $T$.  In the simulations below, we take $T=$\,\SI{50}{ms}, which is sufficient for several measurements of the channel quality in a given antenna. 
We let $\gamma^{(i)}_t$ denote the CQI on link $i$ in time period $t$ if the UE were to receive downlink 
transmissions in that link in that time period.  
For simplicity, we will take $\gamma^{(i)}_t$ as the average SNR on the link.
We assume that the rate follows a practical model \cite{rmmodel,Chen2024_Asilomar_Interpolation,mogensen2007lte}: 
\begin{align}
\label{eq:capacity}
    R^{(i)}_t = \phi_i( \gamma_t^{(i)}) = B_i\min\{ \rho_{\rm max}, \beta \log_2(1 + \gamma_t^{(i)}) \},
\end{align}
where $\phi(\gamma)$ is the CQI-rate mapping function; $B_i$ is the bandwidth of link $i$;
$\beta$ is the coding efficiency with a typical value of 0.6; $\rho_{\rm max}$ is the practical channel capacity subject to the 3GPP MCS table, which is commonly selected from 4.8 - 7.5 based on different QAM strategies. 
The model below can be extended to more complex CQI descriptions and multi-stream transmissions.

As discussed in the Introduction, a critical challenge is that the CQI across all links may not be observable. 
First, two or more antennae may share a common RF receiver chain via a switch implying that only one of those antennas can be used.  Second, even if the two antennas can be activated at the same time, it may not be desirable to turn both of them on to save energy. To model this effect, we let $A_t \subseteq \{1,\ldots,M\}$ denote the set of link indices that were active in period $t$.  The link being active requires that the antenna for that link is switched to an active RF chain 
and that RF chain is tuned to the appropriate frequency of that link.
We assume that the CQI can only be measured on active links.  We describe the observations as:
\begin{align} \label{eq:ot}
    o_t^{(i)}=
    \begin{cases}
    \wh{\gamma}_t^{(i)}, & i\in A_{t},\\
    \bot, & i\notin A_{t},
    \end{cases}
\end{align}
where $\wh{\gamma}_t^{(i)}$ denotes an estimate of the CQI $\gamma_t^{(i)}$ and $\bot$ denotes ``no measurement"
since the link was not active.  
We let $\bs{o}_t = (o_t^{(1)},\ldots,o_t^{(M)})$ denote the vector of observations across links.
Our goal at each time $t$ is to estimate the rates $R^{(i)}_t$ for all links $i = 1, \ldots, M$ based on the past observations $\bs{o}_s$ for $s < t$.  Note that since we are using observations for $s < t$,
our model is predicting the rates one step in the future.

Although not simulated in this work, such rate predictions can be used to guide selection of measurements, cell selection,
and handover decisions.

\section{Neural Network Predictor}

\begin{figure}
    \centering
    \includegraphics[width=1\linewidth]{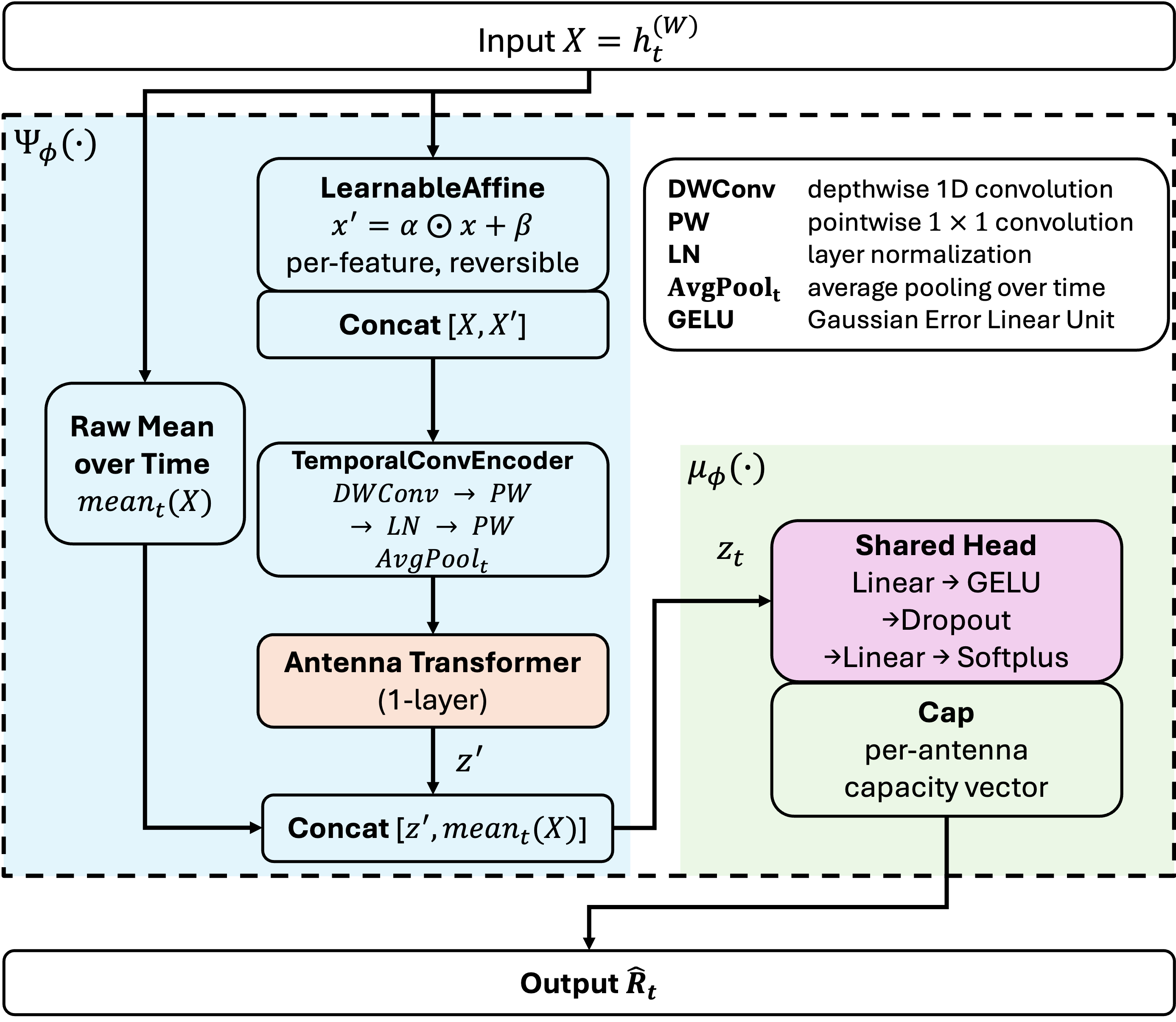}
    \caption{Overall rate estimation network. Each antenna’s recent history is passed through a learnable per-feature affine scaling that preserves units for rate and SNR while improving numerical conditioning. The scaled and raw features are concatenated. A lightweight temporal encoder summarizes each antenna’s $W$-sample window into one embedding. An antenna-level Transformer with one encoder layer mixes cross-antenna dependencies. The result is concatenated with the raw mean over time, which averages only along the time dimension $W$ and keeps the original antenna count $N$ and feature dimension $F$ unchanged. A shared head with Softplus produces non-negative per-antenna rate estimates. A per-antenna capacity cap conditioned on bandwidth and hardware is applied.}
    \label{fig:whole_nn}
\end{figure}

The predictor in \Cref{fig:whole_nn} follows a physically grounded pipeline. At the input, we first transform each observation $o_t^i$ in \eqref{eq:ot} to a 3-dimensional vector:
\begin{equation}
    \bs{y}_t^{(i)} = \begin{cases}
        (1, 10\log_{10}(\wh{\gamma}_t^{(i)}), \phi(\wh{\gamma}_t^{(i)}) & i \in A_t \\
        (0, 0, 0) & i \notin A_t 
    \end{cases}
\end{equation}
where the first component is 1 or 0 indicating if the a measurement on link $i$ was made or not 
(that is, if $i \in A_t$ or $i \notin A_t$). If a measurement was made, the second component is the SNR in dB scale
and the third component is the rate corresponding to the estimated SNR where $\phi(\gamma)$ is the CQI-rate mapping function 
in \eqref{eq:capacity}.
If no measurement was made, the two other components are set to zero.
We then compute a history of the last $W$ such measurements:
\begin{equation}
    \bs{h}_t^{(W)} = \{ \bs{y}_s^{(i)}, ~i=1,\ldots,M, ~ s \in [t-W, t)\}.
\end{equation}
Since each component $\bs{y}_t^{(i)}$ is 3-dimensional, this history window has length $3MW$
where $M$ is the total number of links and $W$ is the window size.

The $W$-step history is first processed by a learnable per-feature affine scaling that preserves units for rate and SNR. We keep both the raw and the scaled versions to retain absolute levels and to ease optimization, then a lightweight temporal encoder compresses the short window into one embedding per antenna. Before prediction we append the raw mean over time computed by averaging only along the time dimension $W$. This operation removes the window axis yet keeps the original antenna count $M$ and feature dimension $F$ unchanged, which preserves absolute scale information useful for calibration. A shared head with Softplus enforces non-negativity and a per-antenna capacity cap reflects bandwidth and hardware limits.

Cross-antenna interaction is modeled on the antenna index rather than on time. We use a one-layer antenna-level Transformer, namely a single encoder block with pre-layer normalization, multi-head self-attention, and a position-wise feed-forward network with residual connections. This mixes dependencies among antennas while keeping capacity controlled for moderate $|\mathcal{I}_{\text{rx}}|$ and limited data. No positional encoding along antennas is required because antenna identities are fixed and are implicitly captured by the learned embeddings.

\section{Numerical Experiments}

\subsection{Experimental Set-up}
The overall procedures for offline dataset generation can be divided into the ray tracing and Data-preparing parts.

\renewcommand{\arraystretch}{1.5}
\begin{table}[h]
\caption{UE mobility pattern}
\label{table:UE_mobility}
\centering
\scriptsize
\begin{threeparttable}
\setlength{\tabcolsep}{3pt}
\begin{tabular*}{\columnwidth}{@{\extracolsep{\fill}} l l c c c @{}}
\hline
\textbf{Component} & \textbf{Parameter [unit]} & \textbf{Low} & \textbf{Medium} & \textbf{High} \\
\hline
\multirow{2}{*}{\textbf{Pedestrian}}
& Max velocity [m/s]           & 2           & 5            & 10 \\
& Heading rate [rad/s]         & $\pm\pi/2$  & $\pm\pi/3$   & $\pm\pi/6$ \\
\hline
\multirow{2}{*}{\textbf{Handset rotations}}
& Yaw increment [rad/50\,ms]   & $\pm\pi/10$ & $\pm2\pi/10$ & $\pm3\pi/10$ \\
& Pitch increment [rad/50\,ms] & $\pm\pi/20$ & $\pm2\pi/20$ & $\pm3\pi/20$ \\
\hline
\end{tabular*}
\begin{tablenotes}[flushleft]
\footnotesize
\item \textit{Note:} Uniform sampling within the listed intervals; ``heading rate'' is pedestrian orientation change.
\end{tablenotes}
\end{threeparttable}
\end{table}

\textbf{Ray tracing:} The generator first applies the UE mobility pattern model list in \Cref{table:UE_mobility} to update the pedestrian trajectory by sampling the movement velocity, acceleration, and heading rate. The UE is navigated within the NYU Tandon campus using a three-level pedestrian mobility model. Each trajectory consists of 1200 time steps, corresponding to 60 s with a step size of 50 ms. The generator uses the generated routes as the UE center positions, and rotates the handset around the body center. This rotation is applied to adjust the position and orientation of each RX relative to the body coordinate frame.

Sionna \cite{sionna}, an open source software that provides powerful ray tracing results based on the open street map, can then simulate the complex downlink channel gain with antenna directivity for multi-path communication between the gNB with different RXs on UE. The capacity and SNR are calculated based on these channels. \Cref{table:simulation} lists the ray tracing simulation parameters. 

\begin{table}[h]
\centering
\caption{Simulation Parameters}
\label{table:simulation}
\begin{tabular}{ll}
\hline
\textbf{Parameter} & \textbf{Value} \\ \hline
\textbf{Carrier frequency (Bandwidth)}       & 15 GHz (200MHz) \\ 
  & 3.5 GHz (100MHz)\\
\hline
\textbf{Tx antenna pattern} & Tr38901 V polarization \\
\hline
\textbf{Transmitter Power} & 51 dBm \\
\hline
\textbf{Noise Power Spectral Density} & -174 dBm/Hz  \\
\hline
\textbf{Noise Figure} & 7 dB  \\
\hline
\textbf{Measure Period} & 50 ms  \\
\hline
\textbf{Code Effeciency $\beta$} & 0.6  \\
\hline
\textbf{Max Spectral Effeciency $\rho_{max}$} & 4.8 bits/Hz  \\
\hline
\textbf{Sionna ray types} & specular, diffuse, refraction  \\
\hline
\textbf{Sionna max interaction depth} & 5 interactions  \\
\hline
\end{tabular}
\label{parameters}
\end{table}
The experiment is conducted in the boundary area of a 15 GHz serving cell, where the channel rates of the two frequency bands frequently exhibit rank reversals. In addition, the self-rotation of the user causes the rates of the RXs operating on the same frequency to alternately become dominant.

\begin{figure}[h]
    \centering
    \includegraphics[width=1\linewidth]{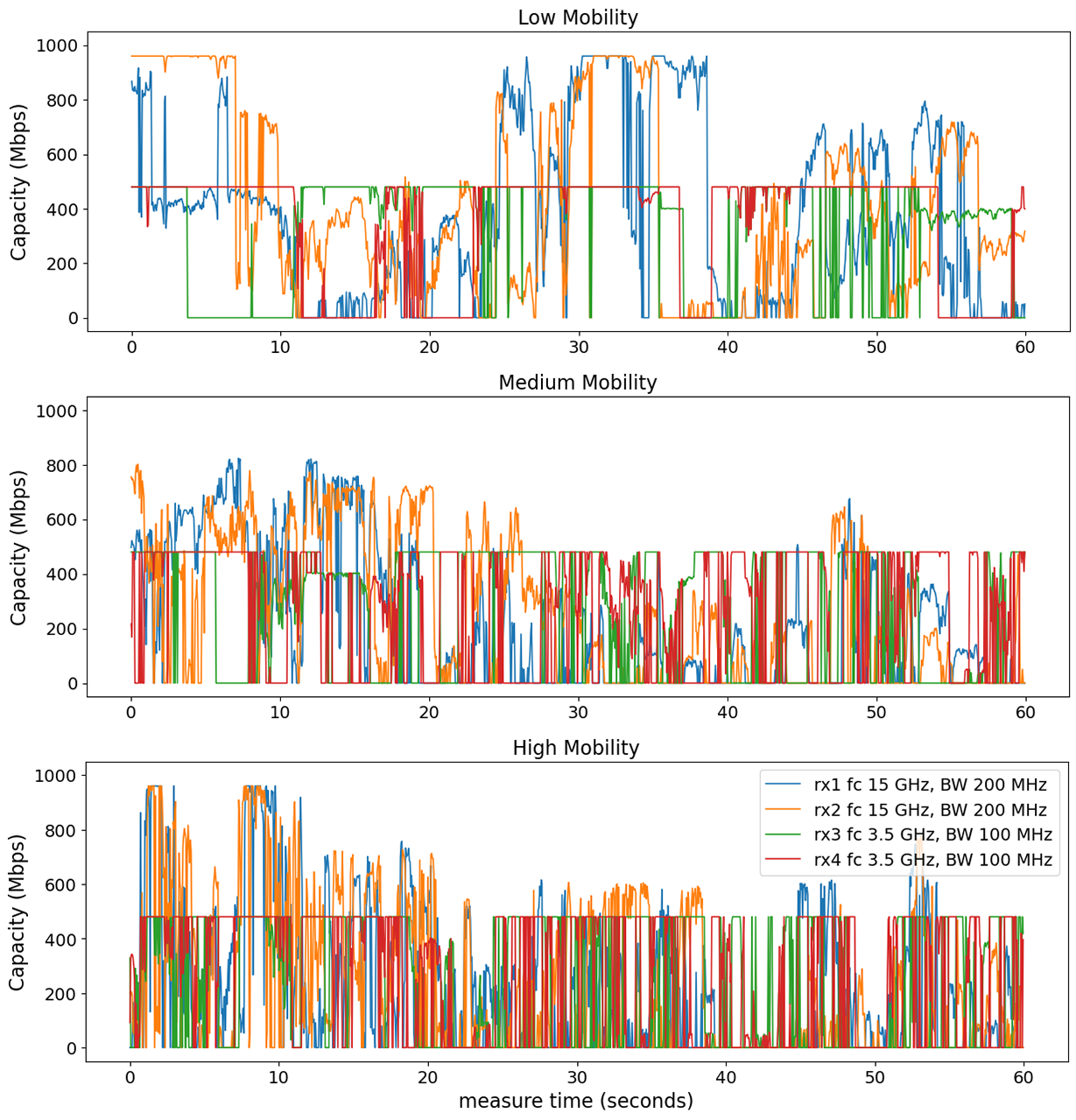}
    \caption{Channel Capacity at three-level mobility. As the mobility level increases, the capacity at each RX exhibits more frequent variations. This increased temporal variability makes it challenging to maintain high transmission rates using only instantaneous feedback. Therefore, accurate rate prediction becomes essential to anticipate channel changes and enable timely adaptation of transmission strategies.}
    \label{fig:low_medium_high}
\end{figure}

For each mobility level, 20 random routes are generated. Each route includes 50 handset rotations, resulting in 1000 instances in total. Each route lasts for 60 seconds, corresponding to 1200 steps (50 ms/step) sampling steps with a step size of 50 ms. Example ray tracing data for each RX is shown in \Cref{fig:low_medium_high}.

\textbf{Data-preparing:} 
Real systems face energy/thermal limits, so not all antennas can be active. To explicitly simulate this missing-measurement regime, we use an adversarial multi-armed bandit policy $\pi_{\mathrm{adv}}$ that defines a probability distribution over subsets,
\[
A_t \sim \pi_{\mathrm{adv}}(\cdot \mid h_t^{(W)}),
\]
 For logging and a baseline, we use an exponential-weights forecaster (EXP3/Hedge) with $\varepsilon$-greedy exploration \cite{Auer02Bandit}. After collecting ray tracing data, we run an $\varepsilon$-greedy exponential-weights bandit ($\varepsilon=0.2$) to choose at each step a hopping observation subset of RX antennas based on historical scores. The chosen subset defines a mask: the UE keeps the corresponding ray-traced capacities and treats the rest as missing. We log the masked observations and the chosen subsets, and use the full capacities as labels to train the rate-only predictor. The generated dataset, including 1000 input instances, is shuffled and divided into 70\%:15\%:15\% for training, validation, and testing. The whole project can be found in \url{https://github.com/Ruibin2000/UE_fc_hopping_v2.git}.

\begin{figure}[t]
    \centering
    \includegraphics[width=\linewidth]{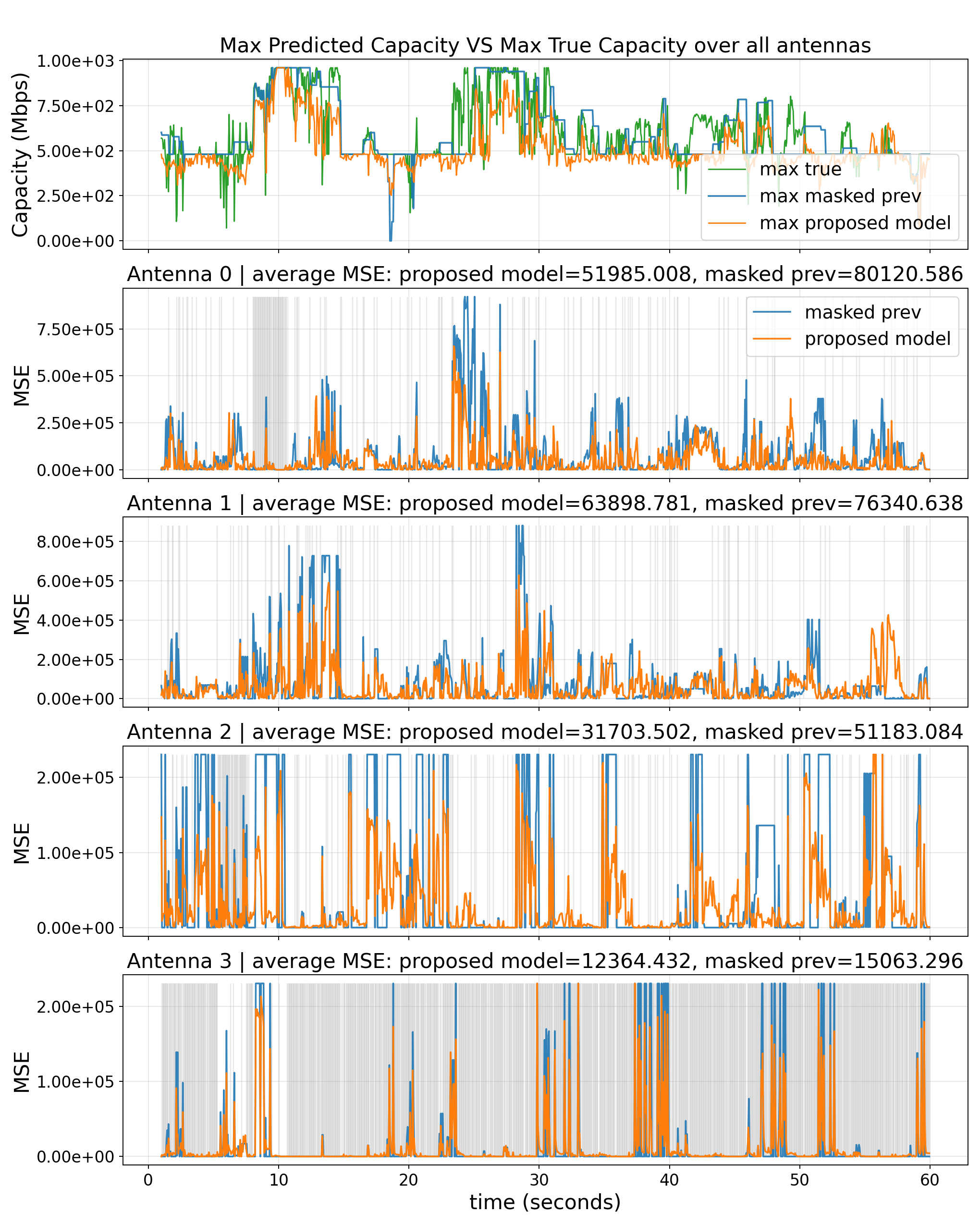}
    \caption{Time-series performance under medium mobility. Top: comparison of the maximum predicted capacity of the proposed model, the masked-previous benchmark, and the maximum ground-truth capacity; the proposed model tracks rapid swings with less lag and fewer overshoots. Bottom: per-antenna MSE over time (gray shading marks timestamps with measurements); the proposed model (orange) is consistently below the masked-previous baseline (blue), indicating lower error both with and without measurements. Overall, per-antenna MSE is reduced by 16--38\% (about 28\% on average).}\label{fig:timeseries_medium}
    \vspace{-4mm}
    \label{fig:time_compare}
\end{figure}
\subsection{Model training}

The transformer-based predictor in \Cref{fig:whole_nn} starts at $t = W$ and is trained on the remaining steps using the Adam optimizer with a learning rate of $10^{-3}$, a dropout rate of $0.2$, and a batch size of 128 for 200 epochs.

\subsection{Benchmarks}
\label{sec:benchmarks}
To evaluate the contribution of the learned predictor, we compare against two simple, predictor-free baselines that rely only on past observations.

\vspace{2pt}\textbf{B1: Full previous observation.}
Always activate all antennas; predict by a one-step lag of the observed per-antenna rates.
\begin{align}
A_t \equiv \mathcal{I}_{\text{rx}},\quad \hat{\mathbf R}_0 := \mathbf R_0,\quad
\hat{\mathbf R}_t := \mu^{\mathrm{full}}(o_{t-1}) := \mathbf R_{t-1},\quad t\ge 1.
\end{align}

\vspace{2pt}\textbf{B2: Bandit-masked previous rate.}
Actions follow the adversarial bandit; an antenna’s prediction is refreshed only if it was active at the previous step, otherwise it is carried forward. After recursion, the first $W$ steps are dropped to align with the predictor’s window.
\begin{align}
A_t \sim \pi_{\mathrm{adv}}(\cdot),\quad \hat{\mathbf R}_0 := \mathbf R_0.
\end{align}
\begin{align}
\begin{split}
\hat{\mathbf R}_t &:= \mu^{\mathrm{mask}}\!\big(\hat{\mathbf R}_{t-1},\,\mathbf R_{t-1},\,\mathbf m_{t-1}\big)\\
&= \mathbf m_{t-1}\!\odot\!\mathbf R_{t-1} + (\mathbf 1-\mathbf m_{t-1})\!\odot\!\hat{\mathbf R}_{t-1},\quad t\ge 1.
\end{split}
\end{align}

\begin{figure}[t]
    \centering
    \includegraphics[width=0.92\linewidth]{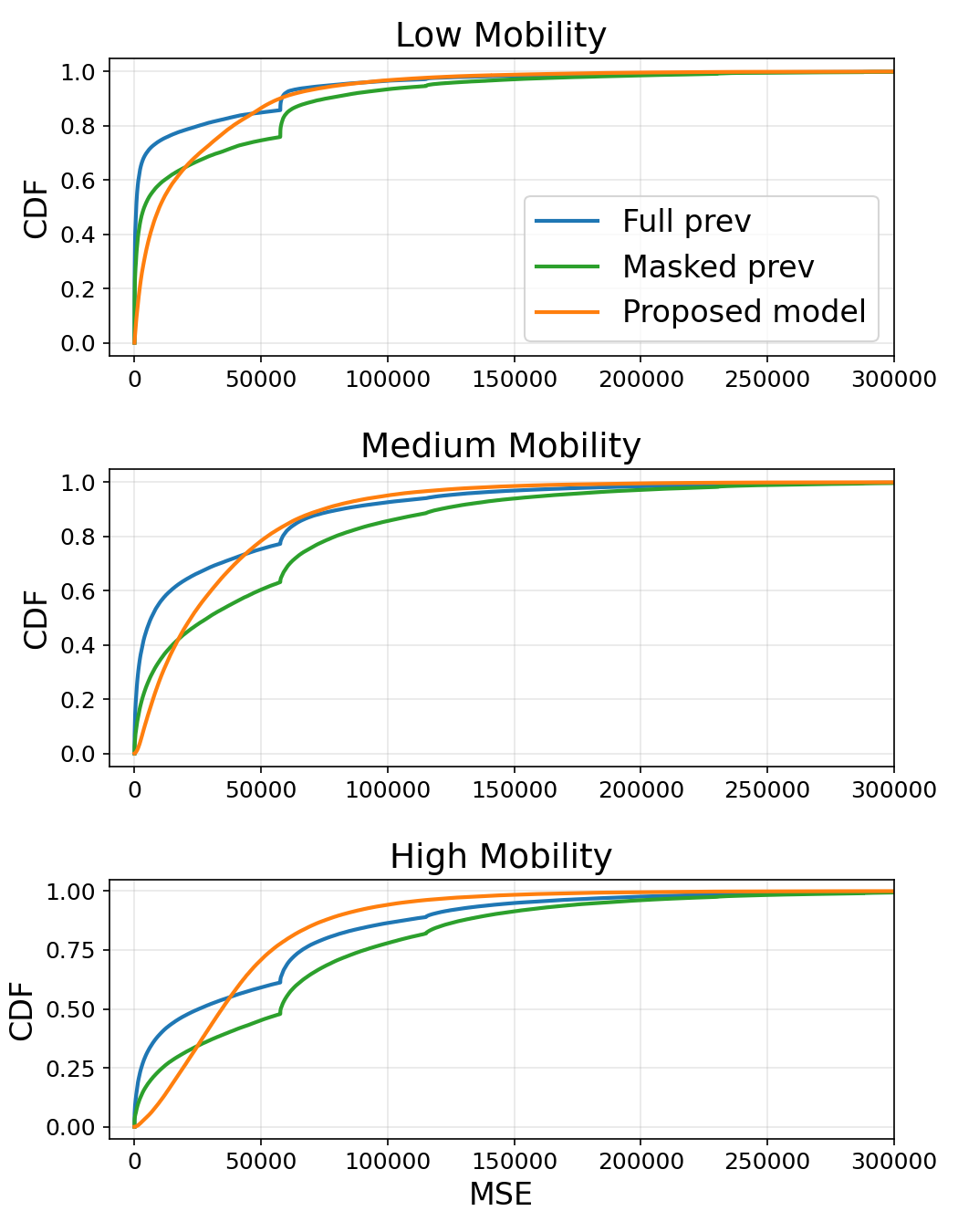}
    \caption{The CDF plots compare the proposed model with the full- and masked-previous baselines under different mobility levels. Although the proposed model has a slightly lower probability of very small errors, it effectively suppresses large errors, leading to a tighter error distribution. This effect becomes more evident as mobility increases, demonstrating improved robustness under rapid channel variations.}
    \label{fig:cdf}
\end{figure}
\subsection{Experiment Results}
\label{sec:exp_results}
\Cref{fig:time_compare} reports time-domain errors for a medium-mobility run. The instantaneous Mean Square Error (MSE) of the proposed predictor (orange) stays below the masked-previous baseline (blue) across most timestamps, yielding lower per-antenna averages with an overall reduction of $\approx$28\%. Gray shading marks measurement times; the gap persists in every antenna, indicating that the predictor exploits temporal structure rather than copying the last observation.

To aggregate performance across runs, \Cref{fig:cdf} presents the Cumulative Distribution Function (CDF) of the MSE over all test data (150 records, 1180 samples each) under low, medium, and high mobility. Although the probability of very small errors is slightly lower, the proposed model markedly suppresses large-error events, producing a left-shifted CDF in the moderate-to-high error range. The advantage strengthens with mobility, demonstrating robustness to rapid channel variation and safer tails for rate selection.

\section{Conclusions}
We formulated a channel tracking problem across multiple bands for a realistic handset with multiple antennas and switching constraints. The tracker uses a transformer-based neural network and, using past measurements, outperforms simple baselines. In this paper, the set of selected antennas is specified. A future step is to consider the \emph{feedback} nature of the problem, where rate predictions are used to select measurements. This setting presents a challenging exploration–exploitation tradeoff and can potentially be explored with reinforcement learning methods.

\bibliographystyle{IEEEtran}
\bibliography{bibl}

@electronic{sionna,
  title        = {Sionna},
  author       = {Hoydis, Jakob and Cammerer, Sebastian and {Ait Aoudia}, Fay{\c{c}}al and
                  Nimier-David, Merlin and Maggi, Lorenzo and Marcus, Guillermo and Vem, Avinash and Keller, Alexander},
  year         = {2022},
  note         = {ver. 1.1.0, [Online]. Available: https://nvlabs.github.io/sionna/}
}

@article{rmmodel,
  author  = {Hu, Yaqi and Yin, Mingsheng and Rangan, Sundeep and Mezzavilla, Marco},
  title   = {Parametrization and Estimation of High-Rank Line-of-Sight {MIMO} Channels With Reflected Paths},
  journal = IEEE_J_WCOM,
  volume  = {23},
  number  = {4},
  pages   = {3808--3822},
  year    = {2024},
  doi     = {10.1109/TWC.2023.3311735}
}

@inproceedings{mogensen2007lte,
  author    = {Mogensen, P. and Na, W. and Kov{\'a}cs, I. Z. and Frederiksen, F. and Pokhariyal, A. and Pedersen, K. I. and Kolding, T. and Hugl, K. and Kuusela, M.},
  title     = {{LTE} Capacity Compared to the Shannon Bound},
  booktitle = {Proc. IEEE Veh. Technol. Conf. (VTC Spring)},
  pages     = {1234--1238},
  year      = {2007}
}

@article{Auer02Bandit,
  author  = {Auer, Peter and Cesa{-}Bianchi, Nicol{\`o} and Freund, Yoav and Schapire, Robert E.},
  title   = {The Nonstochastic Multiarmed Bandit Problem},
  journal = {SIAM J. Comput.},
  volume  = {32},
  number  = {1},
  pages   = {48--77},
  year    = {2002},
  doi     = {10.1137/S0097539701398375}
}

@article{Raghavan18CommMag,
  author  = {Raghavan, V. and Kovacevic, M. and Ratasuk, R. and Ghosh, A. and Rybalko, M. and Subramanian, S. and Hong, S. and Tarighat, A. and Pi, Z.},
  title   = {Spatio-Temporal Impact of Hand and Body Blockage for Millimeter-Wave User Equipment Design at 28 {GHz}},
  journal = {IEEE Commun. Mag.},
  volume  = {56},
  number  = {12},
  pages   = {46--52},
  year    = {2018},
  doi     = {10.1109/MCOM.2018.1800213}
}

@article{Raghavan22TAP,
  author  = {Raghavan, V. and Subramanian, S. and Tarighat, A. and Pi, Z. and Ghosh, A.},
  title   = {Hand and Body Blockage Measurements With Form-Factor User Equipment at 28 {GHz}},
  journal = {IEEE Trans. Antennas Propag.},
  volume  = {70},
  number  = {1},
  pages   = {606--619},
  year    = {2022},
  doi     = {10.1109/TAP.2021.3098537}
}

@inproceedings{Qi17IWCMC,
  author    = {Qi, Wenzhe and Huang, Jie and Sun, Jian and Tan, Yi and Wang, Cheng{-}Xiang and Ge, Xiaohu},
  title     = {Measurements and Modeling of Human Blockage Effects for Multiple Millimeter Wave Bands},
  booktitle = {Proc. IEEE IWCMC},
  pages     = {1604--1609},
  year      = {2017},
  doi       = {10.1109/IWCMC.2017.7986524}
}

@article{Virk20TAP,
  author  = {Virk, Usman Tahir and Haneda, Katsuyuki},
  title   = {Modeling Human Blockage at {5G} Millimeter-Wave Frequencies},
  journal = {IEEE Trans. Antennas Propag.},
  volume  = {68},
  number  = {3},
  pages   = {2256--2266},
  year    = {2020},
  doi     = {10.1109/TAP.2019.2948499}
}

@misc{Shakya24Outdoor,
  author        = {Shakya, D. and Ying, M. and Rappaport, T. S. and Ma, P. and Al-Wazani, I. and Wu, Y. and Wang, Y. and others},
  title         = {Urban Outdoor Propagation Measurements and Channel Models at 6.75 {GHz} {FR1} (C) and 16.95 {GHz} {FR3} Upper Mid-Band Spectrum for {5G} and {6G}},
  year          = {2024},
  eprint        = {2410.17539},
  archivePrefix = {arXiv},
  primaryClass  = {cs.IT},
  note          = {[Online]. Available: https://arxiv.org/abs/2410.17539}
}

@misc{Ying24AngularSpread,
  author = {Ying, M. and Ma, P. and Shakya, D. and Rappaport, T. S.},
  title  = {Angular Spread Statistics for 6.75 {GHz} {FR1} (C) and 16.95 {GHz} {FR3} Mid-Band Frequencies in an Indoor Hotspot Environment},
  year   = {2024},
  note   = {Manuscript, work on angular spread statistics in FR1/FR3 indoor hotspot channels}
}

@article{Mukherjee22Access,
  author  = {Mukherjee, Swagato and Skidmore, Gregory and Chawla, Tarun and Bhardwaj, Anmol and Gentile, Camillo and {\v{S}}eni{\'c}, Jelena},
  title   = {Scalable Modeling of Human Blockage at Millimeter-Wave: A Comparative Analysis of Knife-Edge Diffraction, the Uniform Theory of Diffraction, and Physical Optics Against 60 {GHz} Channel Measurements},
  journal = {IEEE Access},
  volume  = {10},
  pages   = {121720--121731},
  year    = {2022}
}

@article{Fernandes22Access,
  author  = {Fernandes, F. and Duits, M. and van den Biggelaar, J. and Wang, H. and Timmers, M. and Nur, A. and Janssen, G.-J.},
  title   = {Hand Blockage Impact on {5G} mmWave Beam Management Performance},
  journal = {IEEE Access},
  volume  = {10},
  pages   = {106032--106047},
  year    = {2022},
  doi     = {10.1109/ACCESS.2022.3209386}
}

@misc{Ying24RayTracing,
  author = {Ying, M. and Ma, P. and Shakya, D. and Rappaport, T. S.},
  title  = {Site-Specific Location Calibration and Validation of Ray-Tracing Simulator {NYURay} at Upper Mid-Band Frequencies},
  year   = {2024},
  note   = {Preprint, validation of NYURay via site-specific location calibration}
}

@misc{Ju19NYUSIM,
  author        = {Ju, Shihao and Kanhere, Ojas and Xing, Yunchou and Rappaport, Theodore S.},
  title         = {A Millimeter-Wave Channel Simulator {NYUSIM} with Spatial Consistency and Human Blockage},
  year          = {2019},
  eprint        = {1908.09762},
  archivePrefix = {arXiv},
  primaryClass  = {eess.SP},
  note          = {[Online]. Available: https://arxiv.org/abs/1908.09762}
}

@article{Shah22OJCOMS,
  author  = {Shah, Syed Hashim Ali and Rangan, Sundeep},
  title   = {Multi-Cell Multi-Beam Prediction Using Auto-Encoder {LSTM} for mmWave Systems},
  journal = {IEEE Open J. Commun. Soc.},
  volume  = {3},
  pages   = {2194--2209},
  year    = {2022}
}

@article{Alrabeiah20TCOM,
  author  = {Alrabeiah, Muhammad and Alkhateeb, Ahmed},
  title   = {Deep Learning for mmWave Beam and Blockage Prediction Using Sub-6 {GHz} Channels},
  journal = {IEEE Trans. Commun.},
  volume  = {68},
  number  = {9},
  pages   = {5504--5518},
  year    = {2020},
  doi     = {10.1109/TCOMM.2020.3003670}
}

@article{Xu23TWC,
  author  = {Xu, Weihua and Gao, Feifei and Tao, Xiaoming and Zhang, Jianhua and Alkhateeb, Ahmed},
  title   = {Computer Vision Aided mmWave Beam Alignment in {V2X} Communications},
  journal = {IEEE Trans. Wireless Commun.},
  volume  = {22},
  number  = {4},
  pages   = {2699--2714},
  year    = {2023},
  doi     = {10.1109/TWC.2022.3213541}
}

@misc{Yang23Semantics,
  author        = {Yang, Yuwen and Gao, Feifei and Tao, Xiaoming and Liu, Guangyi and Pan, Chengkang},
  title         = {Environment Semantics Aided Wireless Communications: A Case Study of mmWave Beam Prediction and Blockage Prediction},
  year          = {2023},
  eprint        = {2301.05837},
  archivePrefix = {arXiv},
  primaryClass  = {eess.SP},
  note          = {[Online]. Available: https://arxiv.org/abs/2301.05837}
}

@inproceedings{LeiICRA2024,
  author    = {Yin, Mingsheng and Li, Tao and Lei, Haozhe and Hu, Yaqi and Rangan, Sundeep and Zhu, Quanyan},
  title     = {Zero-Shot Wireless Indoor Navigation Through Physics-Informed Reinforcement Learning},
  booktitle = {Proc. 2024 IEEE Int. Conf. Robot. Autom. (ICRA)},
  pages     = {5111--5118},
  year      = {2024},
  doi       = {10.1109/ICRA57147.2024.10611229}
}

@article{Lei2025DTWIN,
  author       = {Li, Tao and Lei, Haozhe and Guo, Hao and Yin, Mingsheng and Hu, Yaqi and Zhu, Quanyan and Rangan, Sundeep},
  title        = {Digital Twin-Enhanced Wireless Indoor Navigation: Achieving Efficient Environment Sensing With Zero-Shot Reinforcement Learning},
  journal      = {IEEE Open J. Commun. Soc.},
  volume       = {6},
  pages        = {2356--2372},
  year         = {2025},
  doi          = {10.1109/OJCOMS.2025.3552277}
}

@inproceedings{Li25RLPhysics,
  author    = {Li, Tao and Lei, Haozhe and Yin, Mingsheng and Hu, Yaqi},
  title     = {Reinforcement Learning With Physics-Informed Symbolic Program Priors for Zero-Shot Wireless Indoor Navigation},
  booktitle = {Proc. Reinforcement Learning Conf. (RLC)},
  year      = {2025}
}

@techreport{3gpp38.101,
  author       = {{3GPP}},
  title        = {{TSG RAN NR; User Equipment (UE) radio transmission and reception; Part 1: Range 1 Standalone (Release 18)}},
  institution  = {3GPP},
  number       = {TS 38.101-1},
  year         = {2024},
  month        = {June},
  note         = {Version 18.3.0},
  url          = {https://www.3gpp.org/DynaReport/38101-1.htm}
}

@techreport{3gpp_release19,
  author       = {{3GPP}},
  title        = {{TSG RAN WG1\#121; Introduction of Rel-19 7-24 GHz channel model enhancements
}},
  institution  = {3GPP},
  number       = {R1-2504964},
  year         = {2025},
  month        = {May},
  note         = {TR38901, CR0026 version 18.0.0},
  url          = {https://www.3gpp.org/ftp/tsg_ran/wg1_rl1/TSGR1_121/Inbox/}
}

@article{kang2024cellular,
  title={Cellular wireless networks in the upper mid-band},
  author={Kang, Seongjoon and Mezzavilla, Marco and Rangan, Sundeep and Madanayake, Arjuna and Venkatakrishnan, Satheesh Bojja and Hellbourg, Gr{\'e}gory and Ghosh, Monisha and Rahmani, Hamed and Dhananjay, Aditya},
  journal={IEEE Open Journal of the Communications Society},
  volume={5},
  pages={2058--2075},
  year={2024},
  publisher={IEEE}
}

@article{testolina2024sharing,
  title={Sharing spectrum and services in the 7--24 ghz upper midband},
  author={Testolina, Paolo and Polese, Michele and Melodia, Tommaso},
  journal={IEEE Communications Magazine},
  volume={62},
  number={8},
  pages={170--177},
  year={2024},
  publisher={IEEE}
}

@inproceedings{shakya2024propagation,
  title={{Propagation measurements and channel models in Indoor Environment at 6.75 GHz FR1 (C) and 16.95 GHz FR3 Upper-mid band Spectrum for 5G and 6G}},
  author={Shakya, Dipankar and Ying, Mingjun and Rappaport, Theodore S and Poddar, Hitesh and Ma, Peijie and Wang, Yanbo and Al-Wazani, Idris},
  booktitle={GLOBECOM 2024-2024 IEEE Global Communications Conference},
  pages={998--1003},
  year={2024},
  organization={IEEE}
}

@article{wang2024review,
  title={Review antenna design for modern mobile phones: A review},
  author={Wang, Yan and Sun, Libin and Du, Zhengwei and Zhang, Zhijun},
  journal={Electromagnetic Science},
  volume={2},
  number={2},
  pages={1--36},
  year={2024},
  publisher={CIE}
}

@misc{Lei2025_FullPosteriorLocalization,
  author = {Lei, Haozhe and Guo, Hao and Svensson, Tommy and Rangan, Sundeep},
  title  = {Beyond Point Estimates: Likelihood-Based Full-Posterior Wireless Localization},
  year   = {2025},
  note   = {arXiv:2509.25719}
}

@inproceedings{Chen2024_Asilomar_Interpolation,
  author    = {Chen, Ruibin and Joy, Jayadev and Hu, Yaqi and Yin, Mingsheng and Mezzavilla, Marco and Rangan, Sundeep},
  title     = {Interpolation Techniques for Fast Channel Estimation in Ray Tracing},
  booktitle = {Proc. Asilomar Conf. Signals, Systems, and Computers},
  pages     = {1383--1388},
  year      = {2024}
}

\end{document}